\documentclass [a4paper,fleqn,12pt]{article}
\usepackage{graphicx}
\usepackage {subfigure,epsfig}

\usepackage {amsmath} \usepackage{amssymb} \usepackage{cite} \usepackage{amsthm}
\numberwithin{equation}{section} \numberwithin{table}{section} \mathindent=0pt
\theoremstyle{plain}

\numberwithin{theorem}{section}

\begin{document}

\title{\textbf{Remarks on rational solutions for the Korteveg - de Vries hierarchy}}

\author{Nikolai A. Kudryashov}

\date{Department of Applied Mathematics\\
Moscow Engineering and Physics Institute\\ (State University)\\
31 Kashirskoe Shosse, 115409, Moscow, \\ Russian Federation}
\maketitle

\begin{abstract}

Differential equations for the special polynomials associated with
the rational solutions of the second Painlev\'e hierarchy are
introduced. It is shown rational solutions of the Korteveg - de
Vries hierarchy can be found taking the Yablonskii - Vorob'ev
polynomials into account. Special polynomials associated with
rational solutions of the Korteveg - de Vries hierarchy are
presented.

\end{abstract}



\section{Introduction}

In this paper our interest is in rational solutions of the Korteveg
- de Vries hierarchy \cite{Lax01}
\begin{equation}\begin{gathered}
\label{1.1}u_{t}+\frac{\partial}{\partial x}\,L_{N+1}[u]=0
\end{gathered}
\end{equation}
Here the operator $L_{N}$ is determined by the Lenard recursion
formula \cite{Lax01, Kudryashov01, Kudryashov02}
\begin{equation}\begin{gathered}
\label{1.2}\frac{d}{dz}\,L_{N+1}[u]=\left(\frac{d^3}{dz^3}+4\,u\,\frac{d}{dz}+
2\,u_{z}\right)\,L_{N}[u],\,\,\quad\,L_{0}[u]=\frac{1}{2}
\end{gathered}\end{equation}
From \eqref{1.2} we have
\begin{equation}\begin{gathered}
\label{1.3}L_{1}[u]=u,
\end{gathered}\end{equation}
\begin{equation}\begin{gathered}
\label{1.4}L_{2}[u]=u_{zz}+3\,u^2,
\end{gathered}\end{equation}
\begin{equation}\begin{gathered}
\label{1.5}L_{3}[u]=u_{zzzz}+10\,u\,u_{zz}+5\,u_{z}^{2}+10\,u^3,
\end{gathered}\end{equation}
\begin{equation}\begin{gathered}
\label{1.6}L_{4}[u]=u_{zzzzzz}+14\,u\,u_{zzzz}+28\,u_{z}\,u_{zzz}+21\,u_{zz}^{3}+\\
+70u^2\,u_{zz}+70\,u\,u_{z}^2+35\,u^4
\end{gathered}\end{equation}
Hierarchy \eqref{1.1} is integrable by the inverse scattering
transform \cite{Kruskal01}. The rational solutions of the Korteveg -
de Vries equation (hierarchy \eqref{1.1} at $N=1$) are studied in
\cite{Ablowitz01, Airault01, Weiss01}.

The aim of this paper is to introduce the hierarchy for the special
polynomials associated with rational solutions of the second
Painlev\'e hierarchy (the Yablonskii - Vorob'ev polynomials
\cite{Yablonskii01, Vorob'ev01, Clarkson01,Clarkson02, Clarkson03,
Demina01}). These special polynomials can be used to find the
rational solutions of the Korteveg - de Vries hierarchy. As
additional result of this paper is special polynomials associated
with rational solutions of the Korteveg - de Vries hierarchy.

\section{Hierarchy for the Yablonskii - Vorob'ev polynomials}

Suppose $u(x,t$ in \eqref{1.1} takes the form
\begin{equation}\begin{gathered}
\label{1.7}u=2\,\frac{\partial\,^{2}\ln{\left(\varphi(t)y(z)\right)}}{\partial
x^2
},\,\,\,\quad\,\,z=\frac{x}{\left((2N+1)t\right)^{\frac{1}{2N+1}}}
\end{gathered}
\end{equation}
where $\varphi(t)$ is arbitrary function of $t$ and $y(z)$ is a new
function of $z$. Then from equation \eqref{1.1} we have the
hierarchy
\begin{equation}\begin{gathered}
\label{1.8}\frac{d\,}{dz}\left(z\,\frac{d \ln
y}{dz}\right)=\frac{1}{2} L_{N+1}\left[2\,\frac{d\,^{2}\ln
y}{dz^{2}}\,\right]
\end{gathered}
\end{equation}

It can be shown that special polynomials $Q^{(N)}_{n}(z)$ associated
with the rational solutions of the second Painlev\'e hierarchy 
satisfy hierarchy \eqref{1.8}.

\begin{table}[h]%[h]
    \caption{The Yablonskii - Vorob'ev polynomials $Q^{(1)}_{n}(z)$} \label{t:1.1}
    \center
       \begin{tabular}[pos]{l}
        \hline\\
        $Q^{(1)}_0(z) = 1$,\\
        $Q^{(1)}_1(z) = z$,\\
        $Q^{(1)}_2(z) = z^3 + 4$,\\
        $Q^{(1)}_3(z) = z^6 + 20z^3 - 80$,\\
        $Q^{(1)}_4(z) = (z^9 + 60z^6 + 11 200)z$,\\
        $Q^{(1)}_5(z) = z^{15} + 140z^{12} + 2800z^9 + 78 400z^6 - 313
        600z^3- 6272 000$,\\
        $Q^{(1)}_6(z) = z^{21} + 280z^{18} + 18480z^{15} + 627 200z^{12}- 17
        248 000z^9 + 1 448 832000z^6$ \\ $\qquad \qquad + 19
        317 760 000z^3 - 38 635 520 000$,\\
        $Q^{(1)}_7(z)=(z^{27} + 504z^{24} + 75 600z^{21} + 5 174 400z^{18} +
        62 092 800z^{15} + 13 039 488 000z^{12}-$\\$\qquad \qquad - 828 731
        904000z^9- 49 723 914 240 000z^6 - 3 093 932 441 600 000)z$ \\ \\
        \hline
    \end{tabular}
\end{table}

\begin{table}[h]%[h]
    \caption{The Yablonskii - Vorob'ev polynomials $Q_{(n)}^{(2)}(z)$} \label{t:1.2}
    \center
       \begin{tabular}[pos]{l}
        \hline \\
        $Q^{(2)}_0(z) = 1$,\\
        $Q^{(2)}_1(z) = z$,\\
        $Q^{(2)}_2(z) = z^3$,\\
        $Q^{(2)}_3(z) = z(z^5-144)$,\\
        $Q^{(2)}_4(z) = z^{10}-1008z^5 -48 384$,\\
        $Q^{(2)}_5(z) = z^{15}- 4032z^{10} - 3 048 192z^5 + 146 313 216$,\\
        $Q^{(2)}_6(z) =z(z^{20}- 12 096z^{15} - 21 337 344z^{10}- 33 798 352
        896z^5 - 4 866 962 817 024)$,\\
        $Q^{(2)}_7(z)=z^3(z^{25}- 30 240z^{20}- 55 883 520z^{15}- 1 182 942
        351 360z^{10}+701 543 488 297 107 456)$\\ \\
        \hline
    \end{tabular}
\end{table}

\begin{table}[h]%[h]
    \caption{The Yablonskii - Vorob'ev polynomials for $Q^{(3)}_{n}(z)$} \label{t:1.3}
    \center
       \begin{tabular}[pos]{l}
        \hline \\
        $Q^{(3)}_0(z) = 1$\\
        $Q^{(3)}_1(z) = z$\\
        $Q_2(z) = z^3$\\
        $Q^{(3)}_3(z) = z^6$\\
        $Q^{(3)}_4(z) = z^3(z^7 + 14 400)$\\
        $Q^{(3)}_5(z) = z(z^{14} + 129 600z^7 - 373 248 000)$\\
        $Q^{(3)}_6(z) =z^{21} + 648 000z^{14}- 24 634 368 000z^7 -35 473 489
        920 000$\\
        $Q^{(3)}_7(z)=z^{28} + 2 376 000z^{21}- 825 251 328 000z^{14}-
        30 436 254 351 360 000z^7
        $\\$\qquad \qquad +43 828 206 265 958 400 000$ \\ \\
     \hline
    \end{tabular}
\end{table}

Assuming $N=1$, $N=2$ and $N=3$  in \eqref{1.8} we have differential
equations for $y(z)\equiv Q^{(1)}_{n}(z)$ in \eqref{1.10} ,
$y(z)\equiv Q^{(2)}_{n}(z)$ in \eqref{1.11} and $y(z)\equiv
Q^{(3)}_{n}(z)$ in \eqref{1.12}
\begin{equation}\begin{gathered}
\label{1.10}yy_{{{
zzzz}}}-4\,y_{{z}}y_{{{zzz}}}+3\,{y_{{{zz}}}}^{2}-y\,y _{{z}}-zy_{{{
zz}}}y+z{y_{{z}}}^{2}=0
\end{gathered}\end{equation}
\begin{equation}\begin{gathered}
\label{1.11}{y}^{2}\,y_{{{ zzzzzz}}}-6\,yy_{{z}}y_{{{zzzzz}}}+5\,yy_
{{{zz}}}y_{{{zzzz}}}+10\,{y_{{z}}}^{2}y_{{{ zzzz}}}-20\,y_{{z}}y_{{{
zz}}}y_{{{zzz}}}+ \\
+10\,{y_{{{zz}}}}^{3}-y_{{z}}{y}^{2}-zy_{ {{
zz}}}{y}^{2}+z{y_{{z}}}^{2}y=0
\end{gathered}\end{equation}

\begin{equation}\begin{gathered}
\label{1.12}{y}^{4}\,y_{{{zzzzzzzz}}}-8\,{y}^{3}\,y_{{z}}\,y_{{{zzzzzzz}}}+28\,{
y}^{2}\,{y_{{z}}}^{2}\,y_{{{zzzzzz}}}+7\,{y}^{3}\,{y_{{{zzzz}}}}^{2}-
\\
-56\,y\,{y_{{z}}}^{3}\,y_{{{zzzzz}}}+112\,y\,{ y_{{z}}}^{2}{y_{{{
zzz}}}}^{2}+28\,{y}^{2}\,{y_{{{ zz}}}}^{2}\,y_{{{
zzzz}}}+28\,y\,{y_{{{ zz}}}}^{4}+
\\
+56\,{y_{{z}}}^{2}{y_{{{zz}}}}^{3}+56\,{y_{{z}}}^{4}y_{ {{
zzzz}}}-112\,{y_{{z}}}^{3}\,y_{{{zz}}}y_{{{ zzz}}}+28\,y{y_
{{z}}}^{2}\,y_{{{zz}}}\,y_{{{zzzz}}}-
\\
-56\,{y}^{2}y_{{z}}y_{{{\it zzz}}}y_{{{
zzzz}}}-112\,y\,y_{{z}}{y_{{{zz}}}}^{2}y_{{{zzz}}
}-y_{{z}}{y}^{4}-zy_{{{zz}}}{y}^{4}+z{y_{{z}}}^{2}{y}^{3}=0
\end{gathered}\end{equation}
Equation \eqref{1.10} for  $Q^{(1)}_{n}(z)$ was found in
\cite{Clarkson01} but equations \eqref{1.11} and \eqref{1.12} for
special polynomials $Q^{(2)}_{n}(z)$ and  $Q^{(3)}_{n}(z)$ are new.
Hierarchy \eqref{1.8} for the special polynomials associated with
rational solutions of the second Painlev\'e hierarchy is new as well
and we need to study this hierarchy in future.

Denote by $w=\frac{d\,\ln y}{dz}$ then equation \eqref{1.8} is
reduced to equation
\begin{equation}\begin{gathered}
\label{1.12a}\,\frac{d\,(z\,w)}{dz}=
L_{N+1}\left[\,\frac{d\,w}{dz}\, \right]
\end{gathered}\end{equation}
Assuming $N=1$, $N=2$  and $N=3$ from \eqref{1.12a} we obtain the
differential equations
\begin{equation}\begin{gathered}
\label{1.13}w_{zzz}+3\,w^{2}_{z}-w-z\,w_{z}=0,
\end{gathered}\end{equation}
\begin{equation}\begin{gathered}
\label{1.14}
w_{zzzzz}+10\,w_{z}\,w_{zzz}+5\,w_{zz}^{2}+10\,w^{3}_{z}-w-z\,w_{z}=0,
\end{gathered}\end{equation}
\begin{equation}\begin{gathered}
\label{1.15}
w_{zzzzzzz}+14\,w_{z}\,w_{zzzzz}+28\,w_{zz}\,w_{zzzz}+21\,w_{zzz}^{3}+
70\,w_{z}^2\,w_{zzz}+\\
+70\,w_{z}\,w_{zz}^2+35\,w_{z}^4-w-z\,w_{z}=0
\end{gathered}\end{equation}
Equation \eqref{1.13} for  $\frac{d \ln(Q^{(1)}_{n}(z))}{dz}$ was
studied in \cite{Clarkson01} but equations \eqref{1.14} and
\eqref{1.15} for special polynomials $\frac{d \ln(
Q^{(2)}_{n}(z))}{dz}$ and $\frac{d \ln(Q^{(3)}_{n}(z))}{dz}$ are
new.

Multiplying \eqref{1.13} on $(2\,w_{zz}-\frac12)$ we have the first
integral of this equation in the form \cite{Clarkson01}
\begin{equation}\begin{gathered}
\label{1.16}w_{{{zz}}}^2-\frac{1}{2}\,w_{{{zz}}}
+2\,(2\,w_{z}^2-\,w)
 \left( w_{{z}}-\frac{z}{4} \right)=C_{1}
\end{gathered}\end{equation}
Taking $C_{1}=\frac{n(n+1)}{4}$ in \eqref{1.16} we have solutions of
\eqref{1.16} in the form of the Yablonskii - Vorob'ev polynomials.
Using $w(z)=\frac{z^2}{8}+p(z)$ in \eqref{1.16} we obtain the
equation in the form \cite{Clarkson01}
\begin{equation}\begin{gathered}
\label{1.17}p_{{{zz}}}^2+4\,\left(p_{z}\right)^3+2\,z\,\left(p_{z}\right)^2\,
-2\,p\,p_{z}-\frac{1}{4}\left(n+\frac12\right)^2=0
\end{gathered}\end{equation}
Hierarchy \eqref{1.12a} is integrable because this is obtained from
the Korteveg - de Vries hierarchy. The study of equation
\eqref{1.13} by the Painlev\'e test \cite{Conte01} yields the Fuchs
indices: $j_1=-1$, $j_2=1$ and $j=6$. These indices correspond to
the arbitrary constants $z_{0}$, $a_{1}$ and $a_{6}$ in the Laurent
series
\begin{equation}\begin{gathered}
\label{1.18}w(z)=\frac{2}{(z-z_{0})}+{a_{1}}+\frac{{z_{0}}}{6}\,\,\left(z-z_{0}\right)-
\left( {\frac {{{z_0}}^{2}}{360}}\,+\frac{{a_1}}{30}\right)
{(z-z_{0}}^{3}-\\
-{\frac {{z_{}}}{72}}\,{(z-z_{0})}^{4}+{a_6}\,{(z-z_{0})}^{5}-
\left({\frac {{{z_{0}}}^{2}}{4320 }}\,+{\frac {{a_1}}{360}}\,
\right){(z-z_{0})}^{6}+...
\end{gathered}\end{equation}

Special polynomials $Q^{(N)}_{n}(z)$ were found in works
\cite{Clarkson02, Demina01}. It is very important to remark that
these polynomials can be found taking into account the differential
- difference hierarchy \cite{Demina02}
\begin{equation}\begin{gathered}
\label{1.18a}Q^{(N)}_{n+1}\,Q^{(N)}_{n-1}\,=z\,(Q^{(N)}_{n})^2\,-2\,(Q^{(N)}_{n})^2\,
L_{N}\left[2\frac{d^2}{dz^2}\left(\ln{Q^{(N)}_{n}}\right)\right],
\,\,\quad\,n\geq1
\end{gathered}\end{equation}

Assuming $Q^{(N)}_{0}(z)=1$ and  $Q^{(N)}_{0}(z)=z$ from
\eqref{1.18a} we have special polynomials $Q^{(N)}_{n}(z)$  at
$n\geq1$. Some of few special polynomials $Q^{(1)}_{n}(z)$
$Q^{(2)}_{n}(z)$ and $Q^{(3)}_{n}(z)$ are given in tables
\eqref{t:1.1}, \eqref{t:1.2} and \eqref{t:1.3}.

\section{Rational solutions of the Korteveg - de Vries hierarchy}

Using the special polynomials associated with the rational solutions
of the second Painlev\'e hierarchy (the Yablonskii - Vorob'ev
polynomials) we can find the rational solutions of the Korteveg - de
Vries hierarchy \eqref{1.1} by the formula
\begin{equation}\begin{gathered}
\label{2.1}u(x,t)=2\,\frac{\partial\,^2\,\ln{Q^{(N)}_{n}(z)}}{\partial
x^2},\,\,\quad\,z=\frac{x}{\left((2N+1)t\right)^{\frac{1}{2N+1}}}
\end{gathered}
\end{equation}

\begin{table}[h]%[t h]
    \center
    \caption{Polynomials $P^{(1)}_n(x,t)$} \label{t:1.4}
    \begin{tabular}{l} %{||c|c|c|c|c|c|c||} %{||c|c|p{65mm}||}
        \hline\\
        $ P^{(1)}_{1}(x,t)=x $, \\
        $ P^{(1)}_{2}(x,t)=x^3+12\,t$, \\
        $ P^{(1)}_{3}(x,t)=x^6+60\,x^3\,t-720\,t^2 $, \\
        $ P^{(1)}_{4}(x,t)=\left(x^9+180\,x^6\,t+302400\,t^3\right)\,x$, \\
        $
        P^{(1)}_{5}(x,t)=x^{15}+420\,x^{12}\,t+25200\,x^9\,t^2+2116800\,x^6\,t^3
        -$\\
        \qquad$-2540166000\,x^3\,t^4- 1524096000\,t^5,$ \\
        $P^{(1)}_{6}(x,t)={x}^{21}+840\,{x}^{18}t+166320\,{x}^{15}{t}^{2}+
        16934400\,{x}^{12}{t}^
        {3}-1397088000\,{x}^{9}{t}^{4}+$\\ \qquad$
        +352066176000\,{x}^{6}{t}^{5}+
        14082647040000\,{x}^{3}{t}^{6}-84495882240000\,{t}^{7}
        $,\\
        $P^{(1)}_{7}(x,t)=x ({x}^{27}+1512\,{x}^{24}t+680400\,{x}^{21}{t}^{2}+139708800\,{
        x}^{18}{t}^{3}+$\\ \qquad$
        +5029516800\,{x}^{15}{t}^{4}+3168595584000\,{x}^{12}{t}^ {5}-
        604145558016000\,{x}^{9}{t}^{6}-$\\$\qquad \qquad-
        108746200442880000\,{x}^{6}{t}^{7}-60897872248012800000\,{t}^{9})$\\
        \\ \hline
    \end{tabular}
\end{table}

\begin{table}[h]%[t h]
    \center
    \caption{Polynomials $P^{(2)}_n(x,t)$} \label{t:1.5}
    \begin{tabular}{l} %{||c|c|c|c|c|c|c||} %{||c|c|p{65mm}||}
        \hline\\
        $ P^{(2)}_{1}(x,t)=x $, \\
        $ P^{(2)}_{2}(x,t)=x^3$, \\
        $ P^{(2)}_{3}(x,t)= x \left( {x}^{5}-720\,t \right) $, \\
        $ P^{(2)}_{4}(x,t)={x}^{10}-5040\,{x}^{5}t-1209600\,{t}^{2}$, \\
        $P^{(2)}_{5}(x,t)={x}^{15}-20160\,{x}^{10}t-76204800\,{x}^{5}{t}^{2}+
        18289152000\,{t}^{3}$,\\
        $P^{(2)}_{6}(x,t)=x \,({x}^{20}-60480\,{x}^{15}t-533433600\,{x}^{10}{t}^{2}-
        4224794112000\,{x}^{5}{t}^{3}-$\\$\qquad \qquad-3041851760640000\,{t}^{4})
        $,\\
        $P^{(2)}_{7}(x,t)={x}^{3} ({x}^{25}-151200\,{x}^{20}t-
        1397088000\,{x}^{15}{t}^{2}-147867793920000\,{x}^{10}{t}^{3}+$\\$\qquad \qquad
        +2192323400928460800000\,{t}^{5})$\\
        \\  \hline\\
    \end{tabular}
\end{table}

\begin{table}[h]%[t h]
    \center
    \caption{Polynomials $P^{(3)}_n(x,t)$} \label{t:1.6}
    \begin{tabular}{l} %{||c|c|c|c|c|c|c||} %{||c|c|p{65mm}||}
        \hline\\
        $ P^{(3)}_{1}(x,t)=x $, \\
        $ P^{(3)}_{2}(x,t)=x^3$, \\
        $ P^{(3)}_{3}(x,t)= x^{6} $, \\
        $ P^{(3)}_{4}(x,t)={x}^{3} \left( {x}^{7}+100800\,t \right) $, \\
        $ P^{(3)}_{5}(x,t)=x \left({x}^{14}+907200\,{x}^{7}t-18289152000\,{t}^{2} \right) $,\\
        $ P^{(3)}_{6}(x,t)={x}^{21}+4536000\,{x}^{14}t-1207084032000\,{x}^{7}{t}^{2}-
        12167407042560000\,{t}^{3}$\\
        $ P^{(3)}_{7}(x,t)={x}^{28}+16632000\,{x}^{21}t-40437315072000\,{x}^{14}{t}^{2}-
        $\\$\qquad \qquad-10439635242516480000\,{x}^{7}{t}^{3}+
        105231523244566118400000\,{t}^{4}$\\
        \\ \hline\\
    \end{tabular}
\end{table}
The four rational solutions of the Korteveg - de Vries equation
(hierarchy \eqref{1.1} at $N=1$) take the form
\begin{equation}\begin{gathered}
\label{3.1}u_{1}(x,t)=-\frac{2}{x^2},
\end{gathered}
\end{equation}
\begin{equation}\begin{gathered}
\label{3.1}u_{2}(x,t)=6\,{\frac {x \left( -{x}^{3}+24\,t \right) }{
\left( {x}^{3}+12\,t \right) ^{2}}},
\end{gathered}
\end{equation}
\begin{equation}\begin{gathered}
\label{3.1}u_{3}(x,t)=-12\,{\frac {x \left(
{x}^{9}+43200\,{t}^{3}+5400\,{x}^{3}{t}^{2} \right) }{ \left(
-{x}^{6}-60\,{x}^{3}t+720\,{t}^{2} \right) ^{2}}},
\end{gathered}
\end{equation}
\begin{equation}\begin{gathered}
\label{3.1}u_{4}(x,t)=-{\frac {20}{{x}^{2}
 \left( {x}^{9}+180\,{x}^{6}t+302400\,{t}^{3} \right) ^{2}}}\,\,
 [{x}^{18}+144\,{x}^{15}t-\\
 -2116800\,{x}^{9}{t}^{3}+22680\,{x}^{12}{t}^{2}-
152409600\,{x}^{6}{t}^{4}+9144576000\,{t}^{6}]
\end{gathered}
\end{equation}

The four rational solutions of the fifth-order Korteveg - de Vries
equation (hierarchy \eqref{1.1} at $N=2$) can be written in the form
\begin{equation}\begin{gathered}
\label{3.1}u_{1}(x,t)=-\frac{2}{x^2},
\end{gathered}
\end{equation}
\begin{equation}\begin{gathered}
\label{3.1}u_{2}(x,t)=-\frac{6}{x^2},
\end{gathered}
\end{equation}
\begin{equation}\begin{gathered}
\label{3.1}u_{3}(x,t)=-{\frac
{12\,({x}^{10}+2160\,{x}^{5}t+86400\,{t}^{2})}{ \left( -{x}^{5}+
720\,t \right) ^{2}{x}^{2}}}
\end{gathered}\end{equation}
\begin{equation}\begin{gathered}
\label{3.1}u_{4}(x,t)=-{\frac {20\,{x}^{3} \left(
{x}^{15}+5040\,{x}^{10}t+23587200\,{x}^{5}
{t}^{2}-12192768000\,{t}^{3} \right) }{ \left(
{x}^{10}-5040\,{x}^{5}t -1209600\,{t}^{2} \right) ^{2}}}
\end{gathered}
\end{equation}
The four rational solutions of the seventh-order Korteveg - de Vries
equation (hierarchy \eqref{1.1} at $N=3$) can be given as the
following
\begin{equation}\begin{gathered}
\label{3.1}u_{1}(x,t)=-\frac{2}{x^2},
\end{gathered}
\end{equation}
\begin{equation}\begin{gathered}
\label{3.1}u_{2}(x,t)=-\frac{6}{x^2},
\end{gathered}
\end{equation}
\begin{equation}\begin{gathered}
\label{3.1}u_{3}(x,t)=-\frac{12}{x^2},
\end{gathered}\end{equation}
\begin{equation}\begin{gathered}
\label{3.1}u_{4}(x,t)=-{\frac
{20\,({x}^{14}-362880\,{x}^{7}t+3048192000\,{t}^{2})}{{x}^{2}
 \left( {x}^{7}+100800\,t \right) ^{2}}}
\end{gathered}
\end{equation}
The rational solutions of the Korteveg - de Vries hierarchy can be
presented in the form of special polynomials taking into account the
arbitrary function $\varphi(t)$ in \eqref{1.7}. These polynomials
are determined by means of formulae
\begin{equation}\begin{gathered}
\label{2.2}P^{(N)}_{n}(x,t)=\left(\,\left(2N\,+1\right)t\,\right)^{\left(\frac{n^{2}+\,n}
{4N+2}\right)}\,Q^{(N)}_{n}(z),
\,\quad\,z=\frac{x}{\left(\,\left(2N+1\,\right)t\,
\right)^{\frac{1}{2N+1}}}
\end{gathered}
\end{equation}
 Using these polynomials we obtain rational
solutions from \eqref{1.11} of the Korteveg - de Vries hierarchy
\eqref{1.1} by formula
\begin{equation}\begin{gathered}
\label{2.3}u^{(N)}_{n}(x,t)=2\,\frac{\partial\,^2\,\ln{P^{(N)}_{n}(x,t)}}{\partial
x^2}
\end{gathered}
\end{equation}
Some of these polynomials $P^{(1)}_{n}(x,t)$, $P^{(2)}_{n}(x,t)$ and
$P^{(3)}_{n}(x,t)$ are given in tables \eqref{t:1.4}, \eqref{t:1.5}
and \eqref{t:1.6}.
\section {Acknowledgments}

This work was supported by the International Science and Technology
Center under Project B 1213.

\end{document}